\documentclass[12pt]{article}
\usepackage[]{epsfig}
\usepackage[centertags]{amsmath}
\usepackage{amsfonts}
\usepackage{amssymb}
\usepackage[english]{babel}
\usepackage{amsmath, amsthm, slashed,url}

\begin{document}

\title{\bf Charged Nielsen-Olesen vortices from a Generalized Abelian Chern-Simons-Higgs Theory }
\author{
Lucas Sourrouille
\\
{\normalsize \it Departamento de F\'\i sica, Universidade Federal do Maranh\~ao,
}\\ {\normalsize\it 65085-580, S\~ao Lu\'is, Maranh\~ao, Brazil}
\\
{\footnotesize lsourrouille@yahoo.es } } \maketitle

\abstract{We consider a generalization of abelian Chern-Simons-Higgs model by introducing a nonstandard kinetic term. In particular we show that the Bogomolnyi equations of the abelian Higgs theory may be obtained, being its solutions Nielsen-Olesen vortices with electric charge. In addition we study the self-duality equations for a generalized non-relativistic Maxwell-Chern-Simons model.  }

\vspace{0.5cm}



\vspace{1cm}
\section{Introduction}
The two dimensional matter field interacting with gauge fields whose dynamics is governed by a Chern-Simons term support soliton solutions\cite{Jk}, \cite{hor}. These models have the particularity to became auto-dual when the self-interactions are suitably chosen\cite{JW,JP}. When this occur the model presents particular mathematical and physics properties, such as the supersymmetric extension of the model\cite{LLW}, and the reduction of the motion equation to first order derivative equation\cite{Bogo}. The Chern-Simons gauge field inherits its dynamics from the matter fields to which it is coupled, so it may be either relativistic\cite{JW} or non-relativistic\cite{JP}. In addition the soliton solutions are of topological and non-topological nature\cite{JLW}.

The addition of non-linear terms to the kinetic part of the Lagrangian has interesting consequences for topological defects, making it possible for defects to arise without a symmetry-breaking potential term \cite{sy}.
In the recent years, theories with nonstandard kinetic term, named $k$-field models, have received much attention. The $k$-field models are mainly in connection with effective cosmological models\cite{APDM} as well as the tachyon matter\cite{12} and the ghost condensates \cite{13}.
The strong gravitational waves\cite{MV} and dark matter\cite{APL}, are also examples of non-canonical fields in cosmology. One interesting aspect to analyze in these models concern to its topological structure. In this context several studies have been conducted, showing that the $k$-theories can support topological soliton solutions both in models of matter as in gauged models\cite{BAi,SG}. These solitons have certain features such as their characteristic size, which are not necessarily those of the standard models. From the theoretical point of view, it has been recently shown, in Ref.\cite{LS}, that the introduction of a nonstandard kinetic term in the non-relativistic Jackiw-Pi model Lagrangian\cite{JP}, leads to a topological model in which the self-duality equations are the same of the relativistic ablelian Chern-Simons-Higgs model.

In this paper we study a Chern-Simon-Higgs model with a generalized dynamics. This nonstandard dynamics is introduced by a function $\omega$, which depend on the Higgs field. We study the Bogomolnyi limit for such system. In particular we will show that choosing a suitable $\omega$, the Bologmolnyi equations of Maxwell-Higgs theory may be obtained. The soliton solutions of these equations are identical in form to the Nielsen-Olesen vortices. The difference lies in the fact that, unlike the usual abelian Higgs model, our vortex solutions have electric charge. Finally, we propose a generalization of a non-relativistic Maxwell-Chern-Simons model introduced by Manton\cite{Manton1}, whose self-duality equation are the Bogomolnyi equations of the Higgs model, and analyze the Bogomolnyi framework, obtaining as a solution the Chern-Simons-Higgs vortices.

\section{Self-dual Chern-Simons solitons}

Let us start by considering briefly the relativistic Chern-Simons-Higgs model and its soliton solutions\cite{JW,JLW}. The dynamics of this model is descried by the action
\begin{eqnarray}
S = \int \,\, d^3 x & \Big( \frac{\kappa}{2}\epsilon^{\mu \nu \rho} A_\mu \partial_\nu A_\rho
+  |D_\mu \phi|^2 - V(|\phi|)  \Big)
\label{Ac1}
\end{eqnarray}
This is $(2+1)$-dimensional model with Chern-Simons gauge field coupled to complex scalar field $\phi(x)$.
Here, the covariant derivative is defined as  $D_{\mu}= \partial_{\mu} + ieA_{\mu}$ $(\mu =0,1,2)$, the metric tensor is  $g^{\mu \nu}=(1,-1,-1)$ and $\epsilon^{\mu\nu\lambda}$ is the totally antisymmetric tensor such that $\epsilon^{012}=1$.

The corresponding field equations are

\begin{eqnarray}
D_\mu D^\mu \phi = -\frac{\partial V}{\partial \phi^*}
\nonumber \\[3mm]
F_{\mu \nu} = \frac{e}{\kappa}\epsilon_{\mu \nu \rho}j^\rho
\label{EqM2}
\end{eqnarray}
where $j^i = \frac{1}{2i}\Big(\phi^* D^\rho \phi - (D^\rho \phi)^* \phi \Big)$. The time component of the second equation in (\ref{EqM2}) is
\begin{eqnarray}
B = \frac{e}{\kappa}j^0\;,
\label{gauss}
\end{eqnarray}
which is the Chern-Simons version of Gauss's law. Integrating this equation, over the entire plane, we obtain the important consequence that any object with charge $Q =e\int d^2 x \rho$ also carries magnetic flux $\Phi = \int B d^2 x$ \cite{Echarge}:

\begin{eqnarray}
\Phi =-\frac{1}{\kappa} Q
\end{eqnarray}

The energy may be found from the energy momentum tensor
\begin{eqnarray}
T_{\mu \nu}= D_\mu \phi^* D_\nu \phi + D_\mu \phi^* D_\nu \phi -g^{\mu \nu}[|D_\alpha \phi|^2 -V(|\phi|)]\;,
\end{eqnarray}
Integration on the time-time component yields
\begin{eqnarray}
E = \int \,\, d^2 x & \Big( |D_0 \phi|^2
+ |D_i \phi|^2 +  V(|\phi|) \Big)
\label{E1}
\end{eqnarray}
In order to find the minimum of the energy, the expression (\ref{E1}) can be rewritten as
\begin{eqnarray}
E = \int \,\, d^2 x & \Big(& |D_0 \phi \mp \frac{i}{\kappa} (|\phi|^2 -\upsilon^2) \phi|^2
+ |( D_1 \pm iD_2)\phi|^2
\nonumber \\ & &
- \frac{1}{\kappa} (|\phi|^2 -\upsilon^2)^2 |\phi|^2 + V(|\phi|) \mp \upsilon^2 B\Big)
\label{E2}
\end{eqnarray}
where we have used the Chern-Simons Gauss law and the identities
\begin{eqnarray}
|D_i \phi|^2 = |( D_1 \pm iD_2)\phi|^2 \mp eB|\phi|^2 \pm \epsilon^{ij} \partial_i J_j
\label{iden}
\end{eqnarray}
and
\begin{eqnarray}
|D_0 \phi \mp \frac{i}{\kappa} (|\phi|^2 -\upsilon^2) \phi|^2 &=&   | D_0 \phi|^2 \pm \frac{i}{\kappa} (|\phi|^2 -\upsilon^2)
[\phi^* D_0 \phi - (D_0 \phi)^* \phi] +
\nonumber \\ & &
\frac{1}{\kappa} (|\phi|^2 -\upsilon^2)^2 |\phi|^2
\label{Id2}
\end{eqnarray}
Thus, if the potential is chosen to take the self-dual form
\begin{eqnarray}
V(|\phi|) = \frac{1}{\kappa} (|\phi|^2 -\upsilon^2)^2 |\phi|^2\;,
\label{PA}
\end{eqnarray}
the expression (\ref{E2}) is reduced to a sum of two squares plus a topological term
\begin{eqnarray}
E = \int \,\, d^2 x & \Big(& |D_0 \phi \mp \frac{i}{\kappa} (|\phi|^2 -\upsilon^2) \phi|^2
+ |( D_1 \pm iD_2)\phi|^2
 \mp \upsilon^2 B\Big)
\label{E32}
\end{eqnarray}
Then the energy is bounded below by a multiple of the magnitude of the magnetic flux (for positive flux we choose the lower signs, and for negative flux we choose the upper signs):
\begin{eqnarray}
 E \geq  \upsilon^2 |\Phi|
 \label{}
\end{eqnarray}
The bound is saturated by solutions to the first-order equations
\begin{eqnarray}
( D_1 \pm iD_2)\phi =0
\nonumber \\
D_0 \phi = \pm \frac{i}{\kappa} (|\phi|^2 -\upsilon^2) \phi
 \label{}
\end{eqnarray}
which, when combined with the Gauss law constraint (\ref{gauss}) become the self-duality equations:
\begin{eqnarray}
( D_1 \pm iD_2)\phi =0
\nonumber \\
B = \pm \frac{2}{\kappa^2} (|\phi|^2 -\upsilon^2) |\phi|^2
 \label{Bogo1}
\end{eqnarray}
These equations are clearly very similar to the self-duality equations of Nielsen Olesen vortices\cite{NO}. However, their solutions are magnetic vortices that carry electric charge and may be topological as well as no topological.

\section{Charged Nielsen-Olesen vortices from abelian Chern-Simons-Higgs model}

We will consider, here, a generalization of the Chern-Simons-Higgs model (\ref{Ac1}). We modify this model by changing the canonical kinetic
term of the scalar field,
\begin{eqnarray}
S = \int \,\, d^3 x & \Big( \frac{\kappa}{2}\epsilon^{\mu \nu \rho} A_\mu \partial_\nu A_\rho
+  \omega(\rho)|D_0 \phi|^2 -|D_i \phi|^2- V(\rho)  \Big)
\label{Acg1}
\end{eqnarray}
The function $\omega(\rho)$ is, in principle, an arbitrary function of the complex scalar field $\phi$ and $V(\rho)$ is the scalar field potential to be determined below. Here, we have abbreviated the notation, naming, $\rho=|\phi|^2$.

Variation of this action yields the field equations

\begin{eqnarray}
&&\frac{\partial\omega(\rho)}{\partial \phi^*}|D_0 \phi|^2 -\omega(\rho)D_0 D^0 \phi + D_i D^i \phi -\frac{\partial V}{\partial \phi^*} =0
\nonumber \\[3mm]
&&B = \frac{e\omega(\rho)}{\kappa}j^0
\nonumber \\[3mm]
&&F_{\mu \nu} = \frac{e}{\kappa}\epsilon_{\mu \nu \rho}j^\rho
\label{geEq}
\end{eqnarray}
The second equation of (\ref{geEq}) is the Gauss's law of Chern-Simons dynamics, modified, here, by the function $\omega(\rho)$.
Notice that $\int \,\, d^2 x e \omega(\rho)j^0$ is the conserved charge associated to the $U(1)$ global symmetry
\begin{eqnarray}
\delta \phi = i\alpha \phi \;,
\end{eqnarray}
Indeed, by the Nother theorem
\begin{eqnarray}
J^{0} = \frac{\partial \cal{L}}{\partial(\partial_0 \phi)}\delta \phi +  \frac{\partial \cal{L}}{\partial(\partial_0 \phi^*)}\delta \phi^* =  \frac{e}{2i}\omega(\rho)\Big(\phi^* D^0 \phi - (D^0 \phi)^* \phi \Big)= e\omega(\rho)j^{0}
\label{Noe2}
\end{eqnarray}
where we have chosen $\alpha=\frac{e}{2}$.

Here, we are interested in time-independent soliton solutions that ensure the finiteness of the action (\ref{Acg1}). These are the stationary points of the energy which for the static field configuration reads
\begin{eqnarray}
E = \int \,\, d^2 x & \Big(-\kappa A_0 B -e^2\omega(\rho)A_0^2 \rho  + |D_i \phi|^2 + V(\rho) \Big)
\label{EJP}
\end{eqnarray}
By varying with respect to $A_0$, we obtain the relation
\begin{eqnarray}
A_0 = -\frac{\kappa}{2e}\frac{B}{\rho \omega(\rho)}
\label{A0}
\end{eqnarray}
Substitution of (\ref{A0}) into (\ref{EJP}) leads to
\begin{eqnarray}
E = \int \,\, d^2 x & \Big( \frac{\kappa^2}{4e^2}\frac{B^2}{\rho \omega(\rho)} + |D_i \phi|^2 + V(\rho) \Big)
\label{EJP1}
\end{eqnarray}
Consider, now, the following choice for the $\omega(\rho)$ function
\begin{eqnarray}
\omega(\rho) = \rho^{-1} \frac{\kappa^2}{2e^2}
\label{omega}
\end{eqnarray}
Then, by using the identity (\ref{iden}), the energy may be written as
\begin{eqnarray}
E = \int \,\, d^2 x & \Big( \frac{1}{2}B^2 + |( D_1 \pm iD_2)\phi|^2 \mp eB\rho + V(\rho) \Big)\;,
\label{EJP1}
\end{eqnarray}
where we have dropped a surface term. So, we have obtained an expression of the energy which is the same of the Abelian Higgs model.
The form of the potential $V(\rho)$ that we choose is motivated by the desire to find self-dual soliton solution.
Thus, if we choose the potential as
\begin{eqnarray}
V(\rho) = \frac{\lambda}{4} (\rho -1)^2\;,
\label{pot}
\end{eqnarray}
the energy may be rewritten as follows
\begin{eqnarray}
E = \int \,\, d^2 x & \Big( \frac{1}{2}[ B \mp e(\rho -1)]^2 +|D_{\pm} \phi|^2 + (\rho -1)^2(\frac{\lambda}{4}-\frac{e^2}{2}) \mp eB \Big)
\label{Ener}
\end{eqnarray}
Notice that the potential (\ref{pot}) is the symmetry breaking potential of the Abelian Higgs model. Also, when the symmetry
breaking coupling constant $\lambda$ is such that
\begin{eqnarray}
\lambda= 2e^2\;,
\end{eqnarray}
i.e. when the self-dual point of the Abelian Higgs model is satisfied, the energy (\ref{Ener}) reduce to a sum of square terms which are bounded  below by a multiple of the magnitude of the magnetic flux:
\begin{eqnarray}
E \geq e |\Phi|
\end{eqnarray}
In order to the energy be finite the covariant derivative must vanish asymptotically. This fixes the  behavior of the gauge field $A_i$ and implies a non-vanishing magnetic flux:
\begin{eqnarray}
\Phi = \int \,\,d^2 x B = \oint_{|x|=\infty} \,\, A_i dx^i  = 2\pi N
\end{eqnarray}
where $N$ is a topological invariant which takes only integer values.
The bound is saturated by fields satisfying the first-order self-duality equations:
\begin{eqnarray}
& &D_{\pm}\phi = ( D_1 \pm iD_2)\phi =0
\\
& &
B= \pm e (\rho -1)
\end{eqnarray}
These are just the Bogomolnyi equations of the abelian Higgs model\cite{Bogo}. The difference lies in the fact that, here, our vortices not only carry magnetic flux, as in the Higgs model, but also $U(1)$ charge. This is a consequence that in our theory the dynamics of gauge field is dictated by a Chern-Simons term instead of a Maxwell term as in Higgs theory. Therefore, as consequence of the Gauss law of (\ref{geEq}), if there is magnetic flux there is also electric charge:
\begin{eqnarray}
Q = e\int \,\, d^2 x \;\; \omega(\rho)j^0 = \kappa\int \,\, d^2 x  \;\;  B = \kappa \Phi
\end{eqnarray}
Since, the fields $A_i$ and $\phi$ satisfy the same self-duality equations as in abelian Maxwell-Higgs theory the solutions will be the same for the same boundary conditions. Another interesting aspect is that, although this is a Chern-Simons-Higgs model, we expect to find only topological solitons in contrast to the usual abelian Chern-Simons-Higgs theory which support both, topological and non-topological solitons. Of course the self-duality equations of ordinary abelian Chern-Simons-Higgs model, and therefore its topological and non-topological solitons, may be achieved by taking $\omega(\rho) =1$.

It is also interesting to compare our study with the results obtained in Ref.\cite{Leenam}. In this last work the authors considered a class of generalization of the abelian Higgs model described by
\begin{eqnarray}
S = \int \,\, d^3 x & \Big( -\frac{1}{4}\omega_1(\rho)F_{\mu \nu}F^{\mu \nu} + |D_\mu \phi|^2 - V(|\phi|)  \Big)
\label{}
\end{eqnarray}
Then, they chose
\begin{eqnarray}
\omega_1(\rho) = \rho^{-1} \frac{\kappa^2}{4e^2}
\end{eqnarray}
and as potential term
\begin{eqnarray}
V(|\phi|)= \frac{e^4}{\kappa^2}\rho(\rho -1)^2\;,
\end{eqnarray}
which is the self-dual potential of the Chern-Simons-Higgs theory.
With this conditions, they were able to obtain the Bogomolnyi equations of the Chern-Simons-Higgs theory
\begin{eqnarray}
( D_1 \pm iD_2)\phi =0
\nonumber \\
B = \pm \frac{2}{\kappa^2} (|\phi|^2 -\upsilon^2) |\phi|^2
 \label{}
\end{eqnarray}
The vortex solutions of this equations are identical in form to the  Chern-Simons-Higgs vortex solutions. However the difference lies in that these solutions have no electric charge, in contrast with the charged vortices of Chern-Simons-Higgs model.
In some way, we have performed an inverse procedure of the developed in Ref.\cite{Leenam}, since we have started from a generalization of abelian Chern-Simons-Higgs theory and we have arrived to the Bogomolnyi equation of the abelian Higgs model, whose solutions are Nielsen-Olesen vortices which possess electric charge. Also it is interesting to note that the dielectric function $\omega_1(\rho)=\rho^{-1}\frac{\kappa^2}{4e^2}$ are very similar to our $\omega(\rho)= \rho^{-1} \frac{\kappa^2}{2e^2}$.

\section{Self-dual soliton solution in a generalized non-relativistic Maxwell-Chern-Simons model }
We conclude this note by analyzing a generalized dynamics of a non-relativistic Maxwell-Chern-Simons model proposed by Manton\cite{Manton1}. This model is governed by a (2+1)-dimensional action consisting on a mixture from the standard Landau-Ginzburg and the Chern-Simons model,
\begin{eqnarray}
S = \int d^3 x \Big(-\frac{1}{2} B^2 + i\gamma ( \phi^\dagger \partial_t \phi + i A_0 |\phi|^2 )
- \frac{1}{2} ( D_i\phi )^\dagger  D_i\phi +\nonumber \\
\kappa (A_0 B + A_2 \partial_0 A_1) + \gamma A_0 + \lambda ( |\phi|^2 -1)^2 -A_i J_i^T \Big)
\label{Ac}
\end{eqnarray}
Here, $\gamma$, $\kappa$ and $\lambda$ are real constants and the term $\gamma A_0$ is related to the possibility of a condensate in the ground state\cite{ByH}. In order to hold the Galilean invariance of the model, the transport current $J_i^T$ should transform as $J_i^T \rightarrow J_i^T + \gamma v_i$ under a boost\cite{Manton1}. Choosing a frame where $J_i^T =0$, the field equations takes the form
\begin{eqnarray}
i\gamma D_0 \phi = -\frac{1}{2} D_i D_i \phi -2\lambda (|\phi|^2 -1) \phi
\nonumber \\[3mm]
\epsilon_{i j} \partial_j B = eJ_i + \kappa \epsilon_{i j} E_j
\nonumber \\[3mm]
\kappa B =  e\gamma (|\phi|^2 -1)
\label{Eq}
\end{eqnarray}
where $E_i =\partial_{i}A_{0}- \partial_{0}A_{i}$ is the electric field.
Using the identity (\ref{iden}) and the Gauss law of (\ref{Eq}) the energy of the model for static field configuration may be written as\cite{HHY}: 
\begin{eqnarray}
 E= \int d^2 x \Big( \frac{1}{2} |(D_1 \pm iD_2)\phi|^2  +(\mp \frac{e^2 \gamma }{2\kappa m} +\frac{e^2 \gamma^2}{2\kappa^2} -\lambda) ( |\phi|^2 -1)^2 \mp \frac{1}{2m} B\Big)
\label{}
\end{eqnarray}
For $\lambda = \mp \frac{e^2\gamma }{2\kappa m} +\frac{e^2\gamma^2}{2\kappa^2}$ the potential terms cancel, and the energy takes the minimum by the fields obeying the first order self-duality equations
\begin{eqnarray}
(D_1 \pm iD_2)\phi =0
\nonumber \\[3mm]
\kappa B =  e\gamma (|\phi|^2 -1)
\label{}
\end{eqnarray}
Notice that for $\kappa =\pm \gamma$ this set of equations become the Bogomolnyi equations of the Higgs model\cite{HHY}.

Let us then propose the following generalization of the Manton model (\ref{Ac})
\begin{eqnarray}
S = \int d^3 x &\Big(&-\frac{\omega(\rho)^{-1}}{2} B^2 + i\gamma \omega(\rho) ( \phi^\dagger \partial_t \phi + i A_0 |\phi|^2 )
- \frac{1}{2} ( D_i\phi )^\dagger  D_i\phi +\nonumber \\
&&\kappa (A_0 B + A_2 \partial_0 A_1) + \gamma \omega(\rho) A_0 - V(\rho)\Big)
\label{Acgen2}
\end{eqnarray}
Here, $V(\rho)$ is again the scalar field potential to be determined below.

By varying with respect to $A_0$ we obtain the Gauss law for this system
\begin{eqnarray}
B=\frac{e\gamma}{\kappa}\omega(\rho)(\rho -1)
\end{eqnarray}
After using the identity (\ref{iden}) and the Gauss law, the energy for the static field configuration may be written as follows:
\begin{eqnarray}
E= \int d^2 x \Big( \frac{e^2 \gamma^2}{2\kappa^2} \omega(\rho)(\rho -1)^2 + \frac{1}{2} |(D_1 \pm iD_2)\phi|^2  \mp \frac{e^2 \gamma }{2\kappa m} \omega(\rho)(\rho -1)\rho + V(\rho)\Big)
\label{ef}
\end{eqnarray}
We will show that the solutions of this theory are related to those present in the abelian Chern-Simons-Higgs model, if we choose a suitable $\omega(\rho)$. That choice is
\begin{eqnarray}
\omega(\rho) = \pm 2\frac{e^2}{\gamma \kappa}\rho
\label{omegaomega}
\end{eqnarray}
Then, the Gauss law takes the form
\begin{eqnarray}
B= \pm 2\frac{e^3}{\kappa^2}\rho(\rho -1)
\end{eqnarray}
and the energy functional (\ref{ef}) is written as
\begin{eqnarray}
E= \int d^2 x \Big( (\rho -1)\rho[\rho(\pm\frac{e^4 \gamma}{\kappa^3} - \frac{e^4}{\kappa^2} + \lambda) + (\mp\frac{e^4 \gamma}{\kappa^3} - \lambda)] + \frac{1}{2} |(D_1 \pm iD_2)\phi|^2  \Big)
\label{ef1}
\end{eqnarray}
where we have chosen a specific form of potential motivated by the desire to find self-dual topological soliton solutions. That form is
\begin{eqnarray}
V(\rho) = \lambda \rho(\rho -1)^2
\end{eqnarray}
In order to the write the expression (\ref{ef1}) as sum of a square term plus a topological term, we choose $\gamma=\kappa$ and $\lambda= \mp \frac{e^4}{\kappa^2} + \frac{e^4}{\kappa^2}$. Then, the energy (\ref{ef1}) becomes
\begin{eqnarray}
E= \int d^2 x \Big( \frac{1}{2} |(D_1 \pm iD_2)\phi|^2 \mp \frac{e}{2} B \Big)
\label{ef2}
\end{eqnarray}
Therefore, the energy is bounded below by a multiple of the magnitude of the magnetic flux, which is saturated by the field satisfying
\begin{eqnarray}
(D_1 \pm iD_2)\phi =0
\nonumber \\[3mm]
B =  \pm 2\frac{e^3}{\kappa^2}\rho(\rho -1)
\label{rqf}
\end{eqnarray}
The equations (\ref{rqf}) are just the set of Bogomolnyi self-duality equations of the abelian Chern-Simons-Higgs model. Thus, starting from a Manton model,in which the self-duality equations are identical to those of the abelian Higgs theory, we were able to construct a generalized model with the same Bogomolnyi equations as those of the abelian Chern-Simons-Higgs model. Another interesting aspect is that in our generalization we have proposed a potential which is six order in the field just like the scalar potential introduced in the  Chern-Simons-Higgs system. It is also interesting to note that the generalized model support also non-topological solitons which are not present in the Manton model.

In summary we have discussed the Bogomolnyi framework for two generalized models obtained by the introduction of a nonstandard kinetic terms in the Lagrangian. In the first case we proposed a Chern-Simons-Higgs generalized system and showed that the self-duality equations are those of the abelian Higgs model. The second case consist on a generalizacion of a non-relativistic Maxwell-Chern-Simons model. We showed, here, that the Bogomolnyi equations of this model are the Chern-Simons Higgs self-duality equations.

\vspace{2cm}
{\bf Acknowledgements}\\
I would like to thank Peter Horvathy for many helpful comments and discussions.  Also I would like to thank Rodolfo Casana and Manoel Messias Ferreira for suggestions and for hospitality during the realization of this work. This work is supported by CAPES(PNPD/2011).

\end{document}